\documentclass[twocolumn,pra,aps,superscriptaddress,showpacs]{revtex4-1}
\usepackage{amssymb}
\usepackage{epsfig}
\usepackage{graphicx}
\usepackage{amsmath}
\usepackage{amsfonts}
\usepackage{hyperref}
\usepackage{natbib}
\begin{document}

\title{Quantum discord and entanglement in Heisenberg XXZ spin chain after quenches}
\thanks{The financial supports from the National Natural Science Foundation of China (Grant Nos.11104021, 11074184 and 11047007) are gratefully acknowledged.}
\author{Ren Jie }
\affiliation{Department of Physics and Jiangsu Laboratory of
Advanced Functional Material, Changshu Institute of Technology,
Changshu 215500, P. R. China}

\author{WU Yin-Zhong } 
\affiliation{Department of Physics and Jiangsu Laboratory of
Advanced Functional Material, Changshu Institute of Technology,
Changshu 215500, P. R. China}

\author{ZHU Shi-Qun }
\affiliation{School of Physical Science and Technology, Soochow
University, Suzhou, Jiangsu 215006, P. R. China}

\begin{abstract}
Using the adaptive time-dependent density-matrix renormalization group method, the dynamics of entanglement and quantum discord of a one-dimensional spin-1/2 XXZ chain is studied when anisotropic interaction quenches are applied at different temperatures. The dynamics of the quantum discord and pairwise entanglement between the nearest qubits shows that the entanglement and quantum discord will first oscillate and then approach to a constant value. The quantum discord can be used to predict the quantum phase transition, while the entanglement cannot.
\end{abstract}


\maketitle

Entanglement is an important resource in quantum information
processing, such as quantum computation and quantum teleportation
\cite{Nielsen,Bennett}. Many researches show that entanglement exists naturally in spin systems at zero temperature. By using the
quantum theory, entanglement can be applied to detect the quantum phase transition points \cite{Osterloh,Osborne,Gu,Amico}, which imply quantum fluctuation due to the change of a parameter in the Hamiltonian at zero temperature. Recently, a different and significant measurement, called quantum discord, was introduced \cite{Ollivier}. It can detect these quantum correlations presented in certain separable mixed states, which cannot be captured by the entanglement of formation. The discord can also be used to predict quantum phase transition points at zero temperature \cite{R,Luo,Sarandy,Bose} and at finite temperatures \cite{Werlang}. It is reported that the measurement of quantum discord can be realized in NMR \cite{Auccaise} and optical experiments \cite{Xu}.

Recently, the progress in manipulating cold-atom systems provides
almost ideal realizations of strongly correlated many-particle
systems in experiments \cite{bloch}. The non-equilibrium physics in
quantum many-body spin systems are attracting much interests of
experiments due to the possible application to spintronics, quantum
information processing, etc. Meanwhile, physicists try
to understand the general aspects of non-equilibrium dynamics in a
comparably simple system. One of the methods is the quench where the system's Hamiltonian is changed suddenly. It corresponds to preparing the system in the ground state of the Hamiltonian $H_I$ at the initial time, and then the Hamiltonian is changed instantaneously to $H_F$ after time $t>0$. The quench dynamics of entanglement and quantum discord in a spin-1/2 transverse XY chain was studied \cite{Sengupta,Nag}. Non-equilibrium dynamics of the XXZ model with quenches has been a very active area of research \cite{Barmettler,Chiara,Ren,Lancaster,Martin}, since such a global interaction quench can be actually realized for atoms in optical lattices \cite{Greiner}. It is interesting to investigate the dynamics of entanglement and quantum discord in Heisenberg spin-1/2 XXZ chain following quenches.

The Hamiltonian of an opened $N$ spin-$1/2$ chain with
the nearest neighbor interaction is given by
\begin{equation}
\label{eq1} H=\sum_{i=1}^{N-1}J(S^x_{i}S^x_{i+1}+S^y_{i}S^y_{i+1}+\Delta
S^z_{i}S^z_{i+1}),\\
\end{equation}
where $S^{\alpha}_i(\alpha=x, y, z)$ are
spin operators on the $i$-th site, and the parameter $N$ is the number of spins
in the chain. The parameter $J$ refers to the ratio of
the nearest-neighbor coupling. For simplicity, $J=1$ is considered. The parameter $\Delta$ denotes the anisotropic interaction. The anisotropic interaction $\Delta>1$ is called N\'{e}el antiferromagnetic phase, and $0<\Delta<1$ is called critical \emph{XY} phase. In this paper, we assume that the anisotropic interaction is time dependent. The initial anisotropic interaction is labeled as $\Delta_I$. After $t>0$, the anisotropic interaction $\Delta_I$ suddenly quenches to $\Delta_F$.

In the Letter, the concurrence is chosen as a measure of the
pairwise entanglement \cite{Wootters}. The concurrence $E$ is
defined as

\begin{equation}
\label{eq2} E = \max \{{\lambda_1 - \lambda_2 - \lambda_3 -
\lambda_4 ,0}\},
\end{equation}
where the quantities $\lambda_i (i=1, 2, 3, 4)$ are the square roots
of the eigenvalues of the operator $\varrho = \rho_{12}(\sigma_1^y
\otimes \sigma_2^y)\rho_{12}^\ast (\sigma_1^y \otimes \sigma_2^y)$ and in descending order. The case of $E=1$ corresponds to the
maximum entanglement between the two qubits, while $E=0$ means that
there is no entanglement between the two qubits.

The quantum discord which measures the amount of quantumness in the state was introduced by \cite{Ollivier},
\begin{equation}
Q(\rho)= I(\rho)-C(\rho),
\label{eq3}
\end{equation}
where $I$  represents the total correlation and $C$ is the classical information.  For an arbitrary bipartite state $\rho_{AB}$, the total correlations, which are expressed by quantum mutual information $I(\rho_{AB})$ in Eq.~(\ref{eq3}), is given by $I(\rho_{AB})=s(\rho_A)+s(\rho_B)-s(\rho_{AB})$, and $s(\rho)=-Tr(\rho\log_2\rho)$, where $\rho_{A(B)}$ is the reduced density matrix obtained from
$\rho_{AB}$ by taking the partial trace over the state space of subsystem
$B(A)$. Quantum conditional entropy can be defined as $s(\rho|\{\hat{B_k}\})=\sum_k p_k s(\rho_k)$.
The final state $\rho_k$ of the composite system, which is the generalization of the classical conditional probability, is given by $\rho_k=\frac{1}{p_k}(\hat{I}\otimes \hat{B_k})\rho (\hat{I}\otimes \hat{B_k})$,
with the probability $p_k={\rm tr}(\hat{I}\otimes \hat{B_k})\rho(\hat{I}\otimes \hat{B_k})$ where $\hat{I}$ is the identity operator of the subsystem $A$. The measurement based quantum mutual information takes the form of
$I(\rho|\{\hat{B_k}\})=s(\rho^A)-s(\rho|\{ \hat{B_k}\})$. The classical information is given by
\begin{equation}
C(\rho)=max _ {\{\hat{B_k}\}}  I(\rho| \{\hat{B_k}\}).
\label{eq8}
\end{equation}
In Eq.~(\ref{eq3}), $Q=0$ means that the measurement has extracted all the information about the correlation between $A$ and $B$ and is classical correlation. While $Q\neq0$ means that the information cannot be extracted by local measurement and the subsystem $A$ gets disturbed in the process, a phenomena not usually expected in classical information theory.

Because of $U(1)$ invariance $ [H,\sum_{tol} S^z]=0$, the expectation values $\langle S^z\rangle=0$ in the ground states of Eq.(\ref{eq1}), and the evolution of the reduced density matrix can be described by
$\rho(t)=\frac 14[{\displaystyle
I_{A,B}+\sum_{k=x,y,z}\langle\sigma_{A}^k\sigma_{B}^k(t)\rangle\sigma_{A}^k\sigma_{B}^k}{\displaystyle]}.
$
The entanglement in the reduced density matrix $\rho(t)$ can be express by \cite{Dowling,Wang}
\begin{equation}
\label{eq10} E(\rho(t))=\frac 12\max\{0,\sum_{k=x,y,z}|\sigma_{A}^k\sigma_{B}^k(t)|-1\},
\end{equation}
and the quantum discord is given by \cite{Mazzola}
\begin{equation}
\begin{array}{cc}
Q(\rho(t))=[g(1-d_x-d_y-d_z)+g(1+d_x-d_y+d_z)+\\
g(1-d_x+d_y+d_z)+g(1+d_x+d_y-d_z)]/4\\
-[g(1+D)+g(1-D)]/2,
\end{array}\label{eq11}
\end{equation}
with $d_k(k=x,y,z)=\langle\sigma_A^k\sigma_B^k(t)\rangle$, $D=\max\{|d_x|,|d_y|,|d_z|\}$, and $g(x)=xlog_2(x)$.

The dynamics of entanglement and quantum discord of a spin-1/2
antiferromagnetic Heisenberg chain is analyzed at zero temperature.
Two kinds of initial states are selected. The first initial state is chosen in \emph{XY} phase region, and the second kind of initial states is in N\'{e}el phase region. We can use exact diagonalization to obtain the ground state for a small system, and then apply size scaling behavior. If the system size is large, the adaptive time-dependent density matrix
renormalization group (t-DMRG) method can be applied with a second-order Trotter expansion of the Hamiltonian \cite{White,Vidal}. In our simulation, a Trotter slicing $dt=0.1$ and the t-DMRG codes with double precision are performed with a truncated Hilbert space of $m = 500$. We calculated a chain of $L=60$ sites with time $t\leq10$ and the relative error is below $10^{-4}$. In the simulation time region, our numerical results are reliable, which have been estimated by keeping trace of the discard weight, and the entanglement entropy increases linearly in time \cite{Chiara,Calabrese}.

\begin{figure}[tb]
\includegraphics[width=0.47\textwidth]{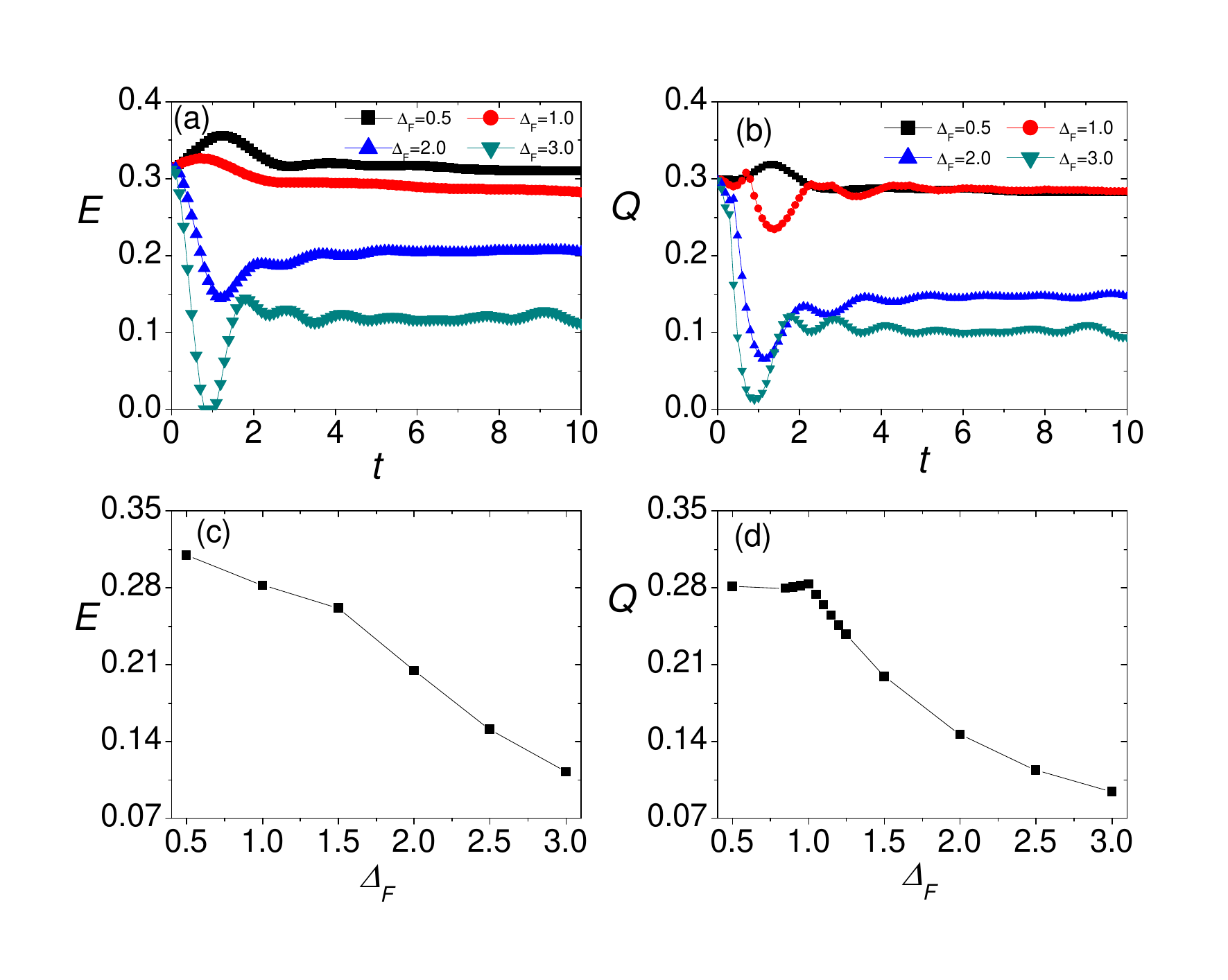}
\caption{\label{fig1} (Color online) The entanglement (a) and quantum discord (b) are plotted as a function of time for different $\Delta_F$, when $\Delta_I=0$ and the temperature of system is zero. The entanglement (c) and quantum discord (d) are plotted as a function of $\Delta_F$ at time $t=10$. Here and in the following all quantities are dimensionless and the dots are t-DMRG data, the lines are fit lines.}
\end{figure}

Firstly, the initial state of Eq. (\ref{eq1}) is in the $\emph{XY}$ phase, then  the anisotropic interaction quench  is applied. The entanglement and quantum discord between the two central qubits are plotted as a function of time for different $\Delta_F$ in Figs. \ref{fig1}(a) and \ref{fig1}(b) respectively for $\Delta_I=0$. In Fig. \ref{fig1}(a), when $\Delta_F=0.5, 1.0$, the entanglement increases firstly with the time increases, then reaches a maximal value, and finally decreases to a constant value. When $\Delta_F=2.0, 3.0$, the entanglement will first decreases with the time $t$ increases, then it drops to a minimal value, and finally reaches to a flat with small oscillations. In Fig. \ref{fig1}(b), the dynamics of quantum discord for $\Delta_F=0.5, 2.0$ is similar to that of entanglement. For $\Delta_F=1.0$, quantum discord has more oscillations in the beginning. To investigate the effects of the anisotropic interaction, the entanglement and discord are plotted as a function of $\Delta_F$ in Figs. \ref{fig1}(c) and \ref{fig1}(d) at the fixed time $t=10$ respectively. In Fig. \ref{fig1}(c), the entanglement decreases with  $\Delta_F$ increases. While in Fig. \ref{fig1}(d), the discord increases slightly when $\Delta_F$ increases, and reaches the maximal value $0.283$ when $\Delta_F=1.0$, and then the discord decreases wihen the interaction $\Delta_F$ increases further. We also investigate the cases of $\Delta_I=0.25, 0.5$, the discord also reaches the maximal value when $\Delta_F=1.0$ at the fixed time.

\begin{figure}[tb]
\includegraphics[width=0.470\textwidth]{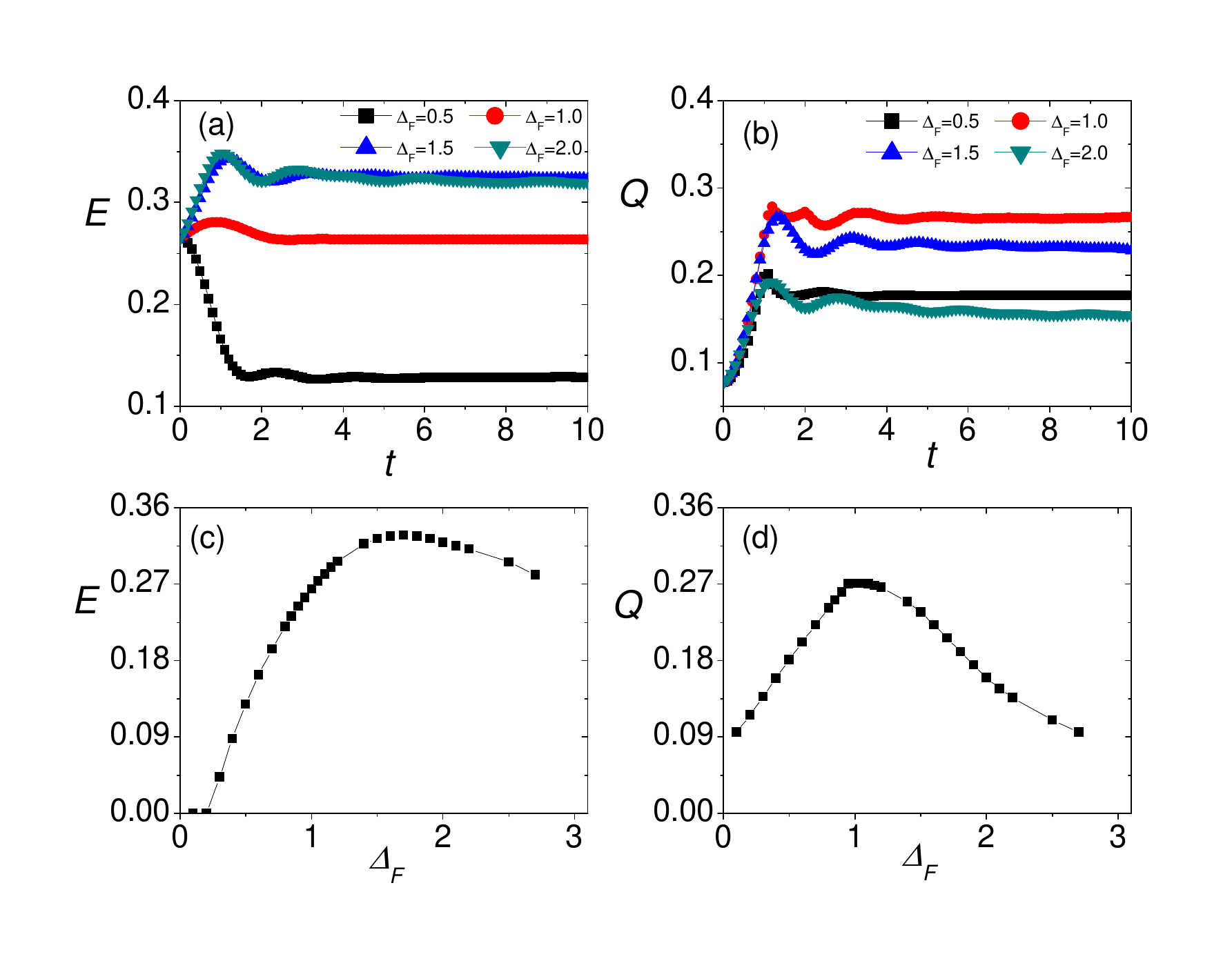}
\caption{\label{fig2} (Color online) The entanglement (a) and quantum discord (b) is plotted as a function of time for different $\Delta_F$, when $\Delta_I=3.0$ and the system temperature is zero. The entanglement (c) and quantum discord (d) is plotted as a function of $\Delta_F$ at the time $t=10$.}
\end{figure}

Secondly, the initial state of Eq. (\ref{eq1}) is in the N\'{e}el phase, then the anisotropic interaction quench is performed. The entanglement and quantum discord between the two central qubits are plotted as a function of time for different $\Delta_F$ in Figs. \ref{fig2}(a) and \ref{fig2}(b) respectively for $\Delta_I=3$. In Fig. \ref{fig2}(a), when $\Delta_F=0.5$, the entanglement decreases with the time increases, then tends to a constant value. For $\Delta_F=1.0, 1.5, 2.0$, the entanglement increases with the time increases, then oscillates to reach a constant value. In Fig. \ref{fig2}(b), the discord increases with the time increases, then oscillates and approaches to a constant value. To find the effects of the anisotropic interaction, the entanglement and discord are plotted as a function of $\Delta_F$ in Figs. \ref{fig2}(c) and \ref{fig2}(d) respectively at a fixed time $t=10$. The entanglement increases when the interaction $\Delta_F$ increases. It reaches the maximal value $0.327$ at $\Delta_F=1.7$, and decreases with $\Delta_F$ increases further. In Fig. \ref{fig2}(d), the discord increases linearly with the interaction increases, and reaches the maximal value $0.271$ at $\Delta_F=1.0$, then decreases with $\Delta_F$ increases. We also study the cases of $\Delta_I=2.0, 2.5$, the discord also reaches the maximal value at $\Delta_F=1.0$ at the same time.

The dynamics of entanglement and quantum discord of a spin-1/2
antiferromagnetic Heisenberg chain is investigated at finite temperature.
It is also started from two kinds of initial states as before. It is known that the density matrix for a system in equilibrium at a finite temperature $T$ can be expressed by
$\label{eq12} \rho=\frac{e^{-\beta H}}{Z}$,
with $\beta=1/kT$ and $Z=Tr(e^{-\beta H})$ is the partition function. The
Boltzmann constant $k$ is set to unity. It is well known that the finite-temperature density matrix renormalization can be calculated by using an enlarged Hilbert space with imaginary time evolution\cite{Feiguin}. , and this method can be extended to study the evolution of the system at finite temperature\cite{Barthel}. A simply method, the time-reversed physical Hamiltonian is applied to reduce the errors \cite{Karrasch}. Here, we adopt this method to study the dynamics of entanglement and discord. In our simulation, a Trotter slicing $dt=0.1,d\beta=0.05$ and the finite temperature t-DMRG codes with double precision are performed with a truncated Hilbert space of $m = 300$. We calculated a chain with $L=30$ sites within $t\leq 5$, and a typically discarded weight can be kept below $\delta w = 10^{-6}$ with relative error below $10^{-4}$.
\begin{figure}[tb]
\includegraphics[width=0.47\textwidth]{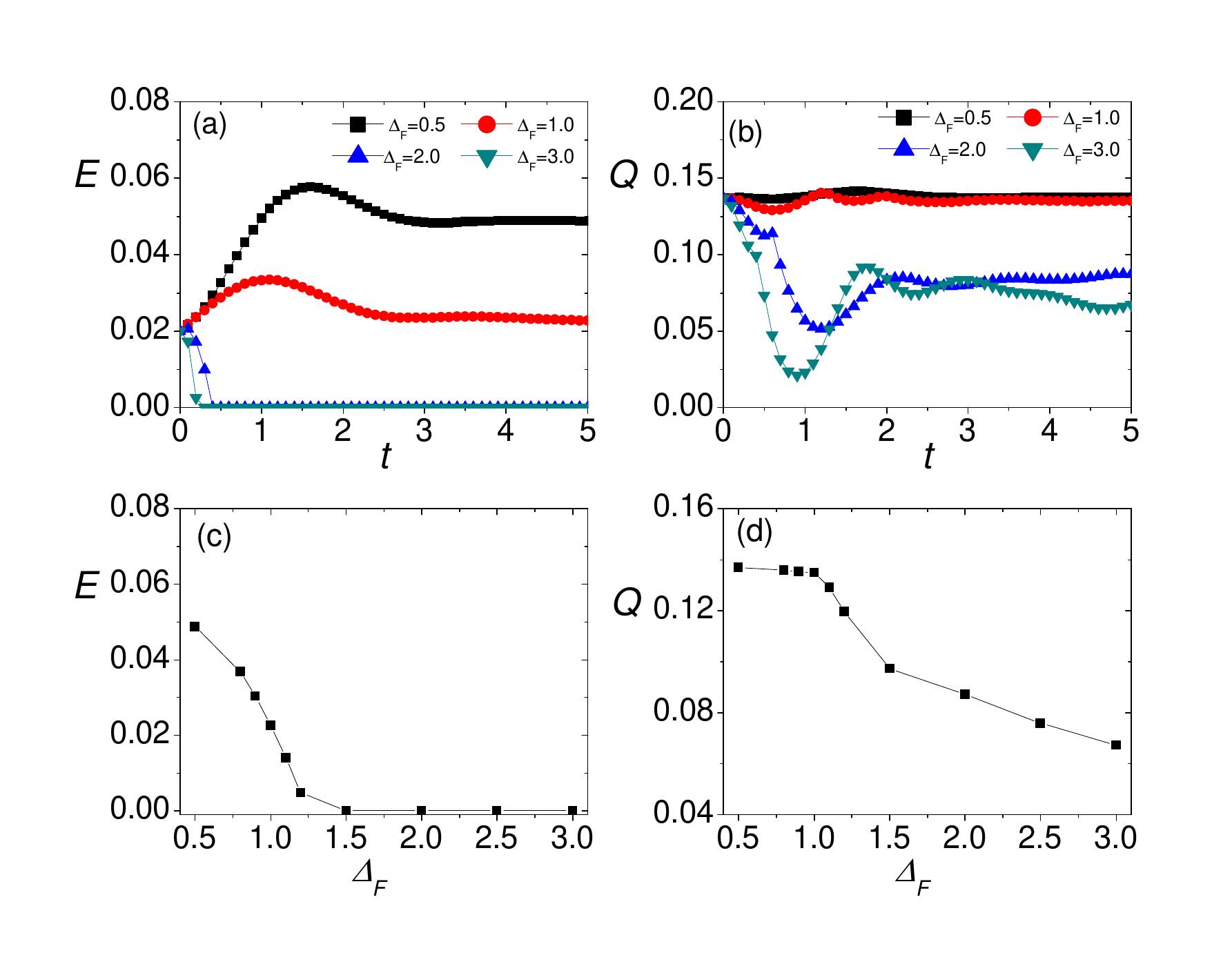}
\caption{\label{fig3} (Color online) The entanglement (a) and quantum discord (b) are plotted as a function of time for different $\Delta_F$, when $\Delta_I=0$ and the system temperature is $T=0.5$. The entanglement (c) and quantum discord (d) are plotted as a function of $\Delta_F$ at the fixed time $t=5$.}
\end{figure}

Firstly, the initial anisotropic interaction is selected as $\Delta_I=0$ and the temperature is chosen as $T=0.5$. The entanglement and quantum discord between the two central qubits are plotted as a function of time for different $\Delta_F$ in Figs. \ref{fig3}(a) and \ref{fig3}(b) respectively. The phenomenon are similar to the case at zero temperature. However, when $\Delta_F=2.0, 3.0$, the entanglement decreases to zero rapidly with time increases. If time is fixed at $t=5$, the entanglement and discord are plotted as a function of the different $\Delta_F$ in Figs. \ref{fig3}(c) and \ref{fig3}(d) respectively. In Fig. \ref{fig3}(c), the entanglement decreases very fast and reaches zero at about $\Delta_F=1.5$ when $\Delta_F$ increases. In Fig. \ref{fig3}(d), the discord decreases very slowly for  $\Delta_F<1.0$. When  $\Delta_F>1.0$, it drops quickly to very small value as $\Delta_F$ increases. Though there is no peak at $\Delta_F=1.0$, the sudden change of the slope at $\Delta_F=1.0$ can be a hint about the quantum phase transition.

\begin{figure}[tb]
\includegraphics[width=0.47\textwidth]{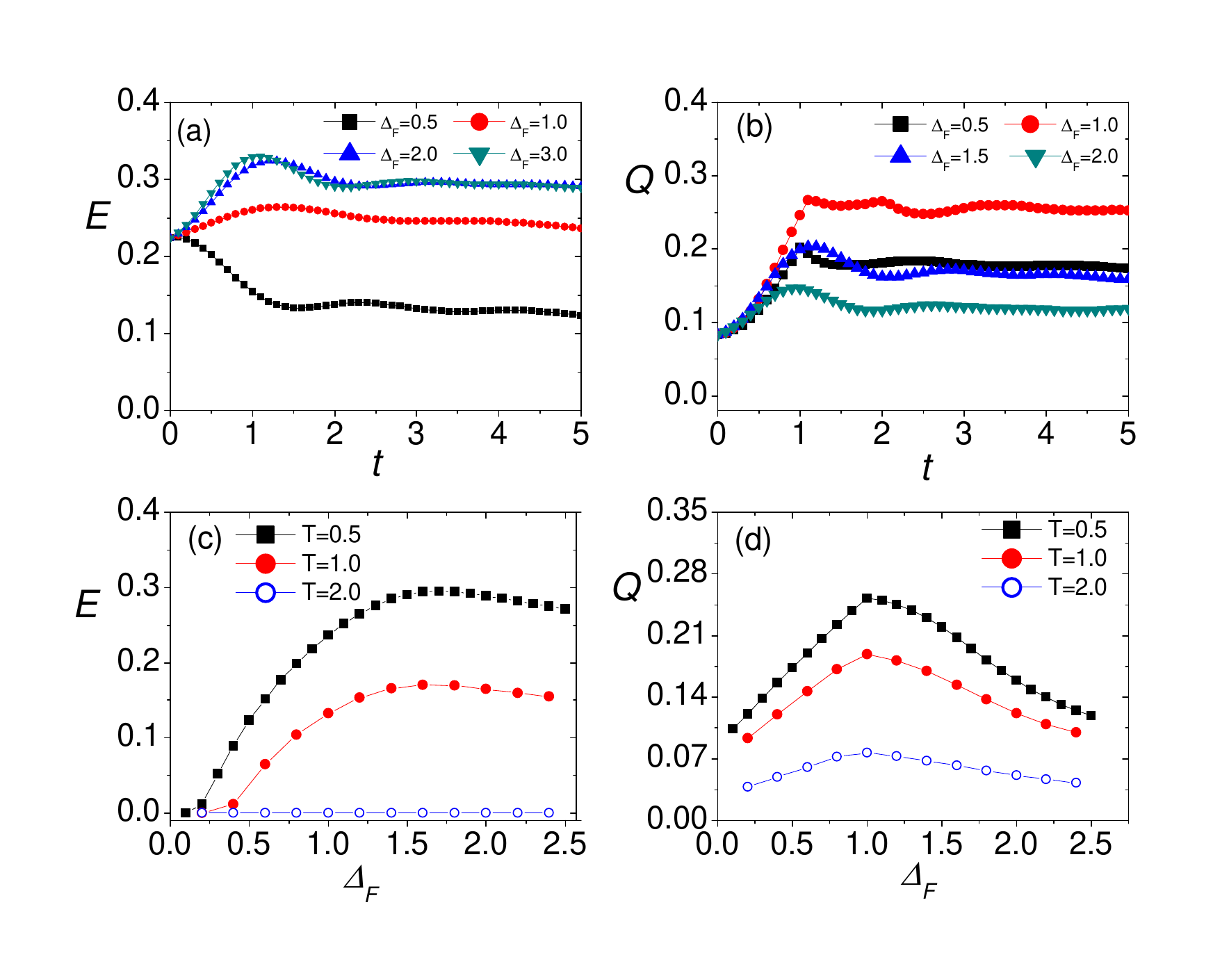}
\caption{\label{fig4}(Color online) The entanglement (a) and quantum discord (b) are plotted as a function of time for different $\Delta_F$, at temperature $T=0.5$ for $\Delta_I=3.0$.  The entanglement (c) and quantum discord (d) are plotted as a function of $\Delta_F$ for different temperature at $t=5$.}
\end{figure}

Secondly, the initial anisotropic interaction is $\Delta_I=3.0$ and the temperature is $T=0.5$.  The entanglement and quantum discord are plotted as a function of time for different $\Delta_F$ in Figs. \ref{fig4}(a) and \ref{fig4}(b) respectively. They are similar to the cases of the quench at zero temperature. The entanglement and discord are plotted as a function of $\Delta_F$ in Figs. \ref{fig4}(c) and \ref{fig4}(d) respectively when the time is fixed. The phenomenon are similar to the case of the quench at zero temperature. The entanglement have a peak located at $\Delta_F=1.7$ for $T=0.5,1.0$, and disappears at $T=2$. The quantum discord have a peak located at $\Delta_F=1.0$ and the peak value decreases with the temperature increases.

At zero temperature, the dynamics of the entanglement shows whether there is a peak in the entanglement with fixed time depends on the initial anisotropic interaction. While for the dynamics of quantum discord, it reaches a maximal value when the final anisotropic interaction is one, irrespective of the initial anisotropic interaction. Therefore this behavior can be employed to predict the quantum phase transition from the critical XY phase to the N\'{e}el antiferromagnetic phase \cite{Werlang}. The entanglement fails to predict the quantum phase transition point while the quantum discord could detect it. At finite temperatures, when the initial anisotropic interaction is smaller than one, the quantum discord only provide a hint of the corresponding quantum phase transition. It is due to the very blunt of the peak of the quench when the temperature is zero. When the temperature increases a little bit, the peak disappears but a sudden change of the slope in quantum discord still exists. When the initial anisotropic interaction is much larger than one, the quantum discord can predict the quantum phase transition, because the peak of the same quench is very sharp at both zero and finite temperatures.

To summarize, the adaptive time-dependent density-matrix renormalization-group method is used to
investigate the dynamics of the entanglement and quantum discord in one-dimensional quantum spin systems when the global anisotropic quenches are performed. The entanglement and discord oscillate to reach a constant value, which depends on the initial and final anisotropic interactions. At zero temperature, whether there is a peak in the entanglement depends on initial anisotropic interaction. While for quantum discord, it reaches a maximal value when the final anisotropic interaction is one, no matter how the initial anisotropic interaction is. This behavior can be used to predict the quantum phase transition points. At finite temperatures, the discord can also be used to predict the quantum phase transition only when the initial anisotropic interaction is greater than one.

\end{document}